\documentclass[12pt,twoside]{article}

\usepackage{asp2006}
\usepackage{natbib}

\begin{document}
\title{The Morphologies and Lifetimes of Transitional Protoplanetary Disks}   
\author{Thayne Currie}   
\affil{NASA-Goddard Space Flight Center}    
\begin{abstract}
I describe new constraints on the lifetimes 
and morphologies of transitional protoplanetary disks from observations of 
1--10 Myr old stars with the \textit{Spitzer Space Telescope}.  New Spitzer results 
clearly show evidence for two kinds of transitional disks and thus two main disk 
evolutionary pathways: disks which form an inner hole/gap and clear from the 
inside out and disks that deplete more homologously.  
Analyzing the disk populations of 1--10 Myr old clusters such as Taurus, IC 348, 
NGC 2362, and $\eta$ Cha show that the mean transitional disk lifetime must be 
an appreciable fraction of the mean protoplanetary disk lifetime: $\approx$  
1 Myr out of 3--5 Myr.  
The varieties of transitional disk SEDs and correlations 
with other disk diagnostics are consistent with multiple mechanisms responsible 
for clearing disks.
\end{abstract}
\section{Background}   
\textit{Transitional} protoplanetary disks bridge the evolutionary gap 
between luminous optically-thick \textit{primordial} disks of gas and small 
dust, which presumably have yet to make planet-mass bodies and gas-poor/free 
optically-thin \textit{debris} disks, which have ended any gas giant planet formation
 \citep{Strom1989, CurrieLada2009}.  
Stars surrounded by transitional disks have near-to-mid IR dust emission intermediate
between primordial disk-bearing stars and diskless photospheres,
implying that much of their solid mass is in the process of being lost 
from the system and/or incorporated into large planetesimals/protoplanets.  
Ground-based studies of transitional disks from IR to submm photometry 
find evidence for structural features in these disks, large inner holes/gaps in the disks' dust
distribution, indicative of active disk dispersal \citep[e.g.][]{Calvet2002}.
Thus, transitional disks may provide valuable insights into how and when 
planet formation ends.

The unprecedented mid-IR sensitivity of the \textit{Spitzer Space Telescope}
made spectroscopic observations of nearby transitional disks and 
photometry for many transitional disks beyond $\sim$ 200--400 pc 
accessible for the first time.  In this contribution, I summarize new Spitzer 
results that clarify our understanding of morphologies and lifetimes of 
transitional disks.  These results reveal
two types of transitional disks, which may be evidence for a range of 
mechanisms responsible for dispersing disks and show that 
the transitional disk phase typically comprises an appreciable fraction of the total 
protoplanetary disk lifetime.  Furthermore, analyzing the accretion frequency/rate and submillimeter 
fluxes of transitional disks provides some insight into the processes 
that are plausibly responsible for transitional disk morphologies and thus 
potentially crucial for dispersing disks and shutting off planet formation.
\section{Spitzer Results on the Morphologies of Transitional Disks}    
Like primordial disks, transitional disks are considered to be a particular kind of 
 "first-generation", protoplanetary disk.   Unlike primordial disks, they show strong 
evidence of dust removal.  While some early papers discussing transitional disks 
differed about how this dust removal proceeds\citep[e.g.][]{Strom1989,Skrutskie1990}, 
pre-Spitzer studies typically associated the term "transitional disk" with "disks with 
inner holes".  Such disks show the most dramatic evidence for dust 
clearing, clearing in such disks could be confirmed by spatially-resolved imaging, 
and physical processes responsible for disk dispersal often lead to an inside-out 
clearing \citep[e.g.][]{Clarke2001, Quillen2004}.   New Spitzer results 
better characterize the properties of transitional disks with inner holes 
and also find that transitional disks may also have a second morphology, exhibiting
 a more global or 'homologous' depletion of dust.   
  Below, we describe properties of both transitional disk morphologies in more detail.

\begin{figure}[h]
\epsscale{0.65}
\plotone{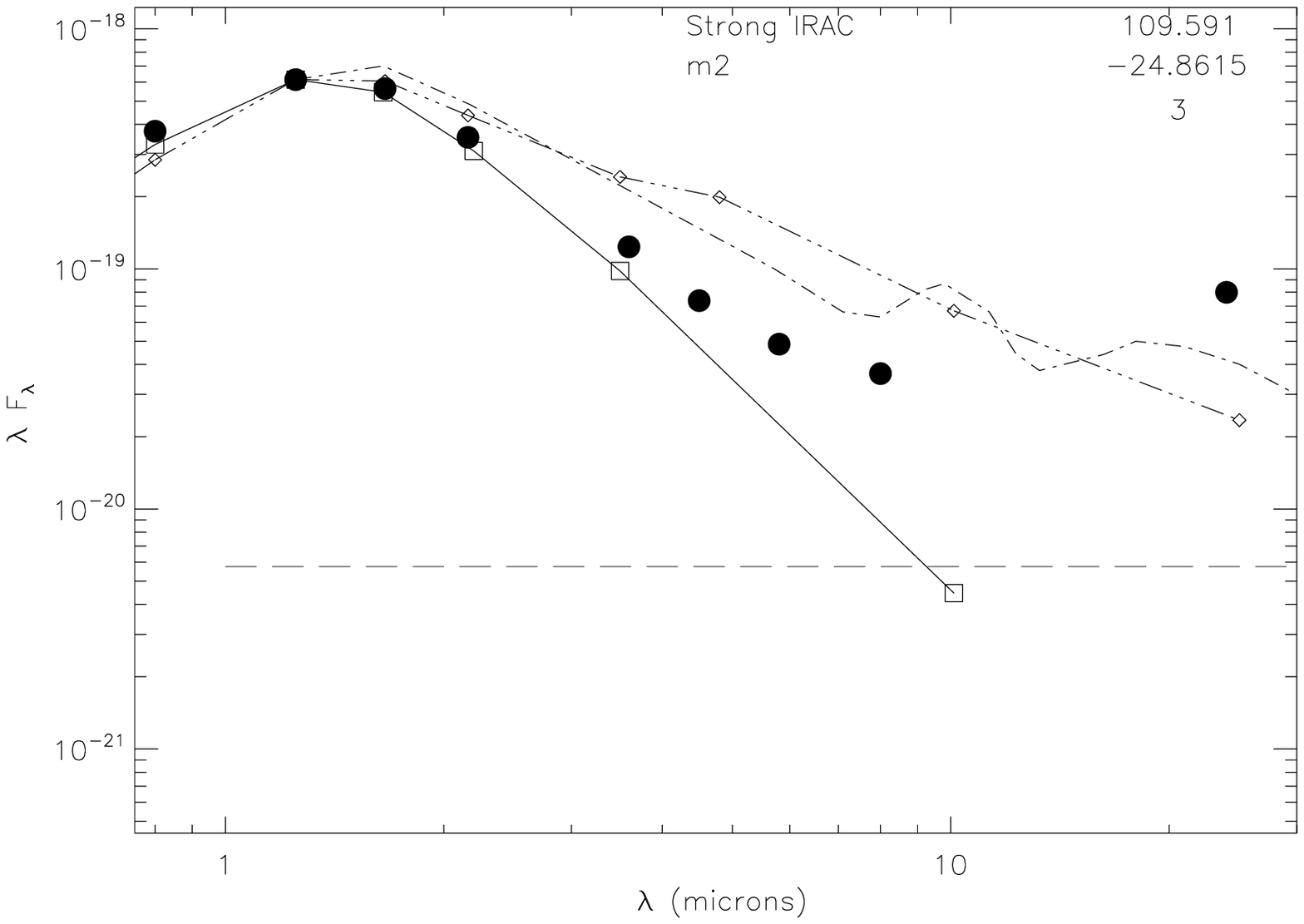}
\plotone{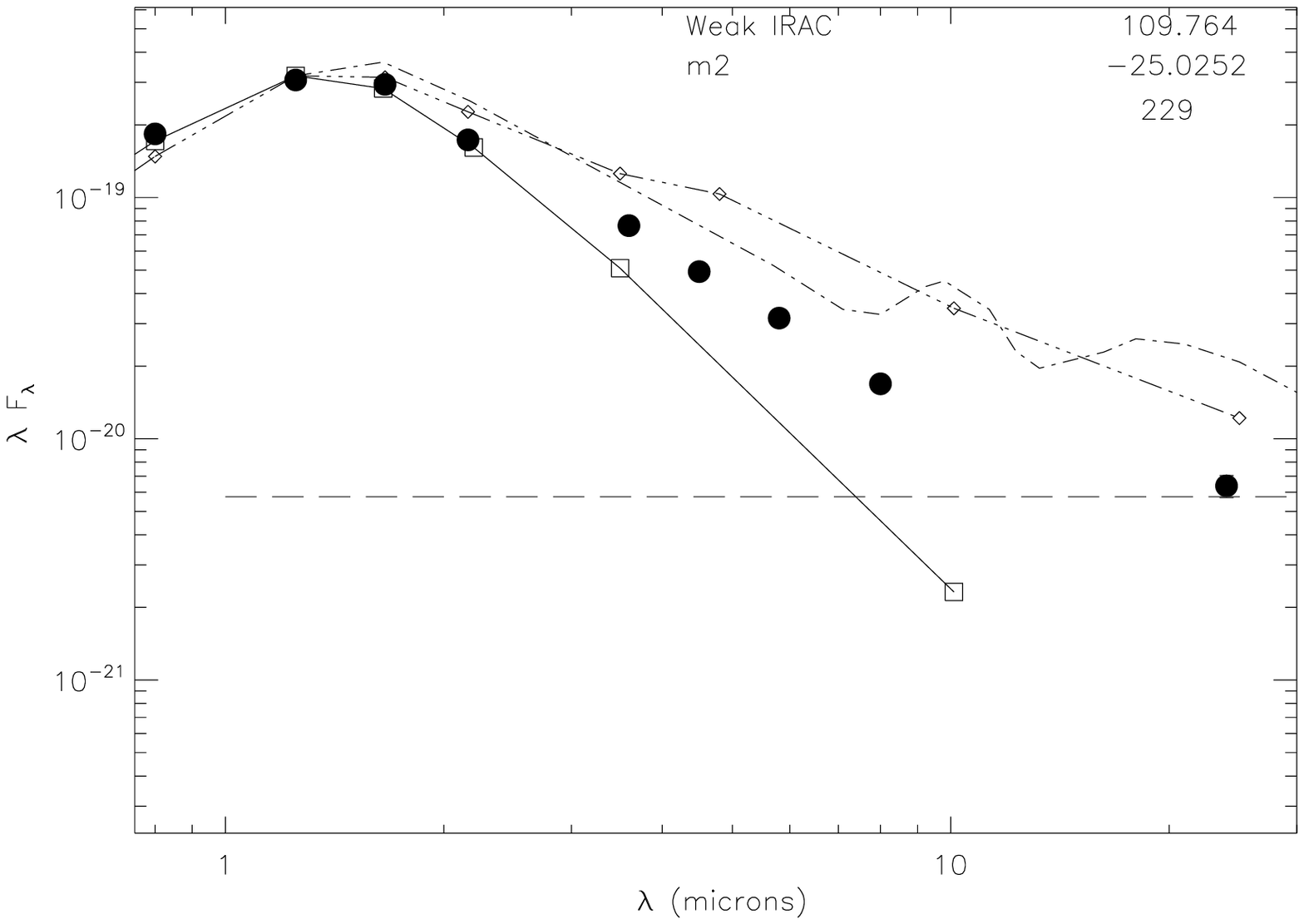}
\caption{SEDs for a transitional disk with an inner hole (top) and a homologously depleted 
transitional disk (bottom) in NGC 2362 compared to the lower-quartile Taurus SED (dash-dot) and flat 
disk SED (dash-three dots/diamonds).}
\end{figure}
\begin{itemize}
\item \textbf{Transitional disks with inner holes/gaps} have photospheric or 
weak, optically-thin 5--10 $\mu m$ emission unlike primordial disks but have more optically-thick emission
at longer wavelengths (e.g. 10--20 $\mu m$) like primordial disks.  Because longer wavelength emission 
preferentially originates from colder, more distant regions of a disk, their SEDs provide 
evidence for a substantial depletion of warm dust with respect to their cold dust population.  
The strong radially-dependent clearing of dust in these disks implies an inside-out dispersal 
of dust.

The presence of an inner hole is consistent with sophisticated SED modeling of photometric and 
spectroscopic data \citep[e.g.][]{Calvet2005} and confirmed by spatially-resolved submillimeter observations \citep[e.g.][]{Hughes2008}.
A slight variant on this morphology is a disk whose weak mid-IR emission provides evidence for 
large disk regions with optically-thin warm dust and an optically-thick outer disk made of cold 
dust but whose near-IR emission reveals optically-thick hot dust close to the star.   
The SEDs of these disks are then more consistent with a large gap instead of an inner hole 
extending to the stellar magnetosphere \citep{Espaillat2009}.
Figure 1 shows the SED for a M2 star in NGC 2362 that has a transitional disk with a $>$ 4 AU inner hole \citep{CurrieLada2009}. 

Typical inner hole sizes inferred from SED modeling of Spitzer IRS spectra range from 
$\sim$ 4 AU to $\sim$ 60 AU \citep[e.g.][]{Calvet2005,Kim2009}.  Broadband IR photometry 
alone may identify transitional disks with large inner holes/gaps around solar-mass stars, where the 
star-to-disk contrast is good \citep{Ercolano2009}.  However, the IR fluxes 
for other transitional disks with inner holes/gaps can be very similar to those for primordial disks 
\citep[e.g. SZ Cha,][]{Kim2009}.  Moreover, even Spitzer IRS spectroscopy may fail to distinguish between 
primordial disks and transitional disks with smaller, $\sim$ AU-scale cleared regions, especially if 
they also have optically-thick near-IR emission (C. Espaillat, pvt. comm.).
Thus, many transitional disks with smaller holes/gaps probably have yet to be discovered.
Analysis of these systems would probe the earliest stages in inside-out disk clearing.

\item \textbf{Homologously depleted transitional disks} lack evidence for a 
strong radially-dependent depletion of dust and, therefore, an inner hole.  
  However, their mid-IR fluxes are substantially weaker than typical primordial disk fluxes, 
lying below the lower-quartile Taurus SED \citep[][]{CurrieLada2009}.  Their emission 
is also weaker than that for geometrically flat, reprocessing blackbody disks viewed face on 
\citep[e.g.][ see Figure 1]{KenyonHartmann1987}.  Because face-on flat disks exhibit the 
weakest IR emission that can come from an optically-thick disk\footnote{Disks that are viewed 
close to edge on and disks with flaring have stronger emission relative to the stellar photosphere 
\citep[e.g.][]{Chiang1999, KenyonHartmann1987, Wood2002}.}, homologously depleted disks 
have a substantially reduced IR optical depth and thus a depleted mass of IR-emitting dust.  
If this IR emitting dust traces the bulk disk mass, then these disks are losing disk mass 
simultaneously over a wide range of radii.

Disk masses for homologously depleted disks, as inferred from submm fluxes and from modeling IR-to-submm SEDs, 
are systematically lower than those for primordial disks \citep[M$_{d}$ $<$ 10$^{-3}$ 
M$_{\odot}$,][]{CurrieLada2009,Cieza2010}.  Additionally, these disks have a lower frequency 
of accretion and lower typical accretion rates \citep{Muzerolle2010, Cieza2010}; disks lacking 
evidence for accretion have the lowest masses.  In light of these properties,  
 homologously depleted disks are much more likely undergoing an overall depletion of disk 
mass, not simply dust settling.  Thus, homologously depleted disks likely represent a distinct 
evolutionary outcome \citep[][]{Muzerolle2010, Cieza2010,CurrieLada2009}.
\end{itemize}
\section{Spitzer Results on the Lifetimes of Transitional Disks}
Comparing the frequencies of transitional disks to primordial disks derived from SED modeling 
for 1--10 Myr-old clusters constrains the mean transitional disk lifetime.  Traditionally, 
the transitional disk lifetime was thought to be short, $\sim$ 0.01--0.1 Myr, based on 
the paucity of transitional disks in Myr-old star-forming regions like Taurus.  
However, if disks spend $\approx$ 0.01--0.1 Myr of their 3--5 Myr lifetime as transitional disks,
 then transitional disks should also be far rarer than primordial disks in older (3--10 Myr) clusters.  

Spitzer observations show that $<$ 20\% of disks surrounding 1--2 Myr-old 0.5--1.4 M$_{\odot}$ stars are 
in transition, but at least 50\% and as many as 80\% of disks are transitional disks by $\sim$ 5--6 Myr 
\citep{CurrieLada2009, SiciliaAguilar2009}.  Based on these results, \citet{CurrieLada2009} show that the transition timescale  
 is $\approx$ 1 Myr.  To investigate this issue further, I have applied the disk classification scheme from 
\citet{CurrieLada2009} to other 1--10 Myr-old clusters, modified to include the flat reprocessing disk SED as 
an additional check on whether a disk is primordial or whether it is a homologously depleted transitional disk.  

\begin{figure}[h]
\centering
\epsscale{0.7}
\plotone{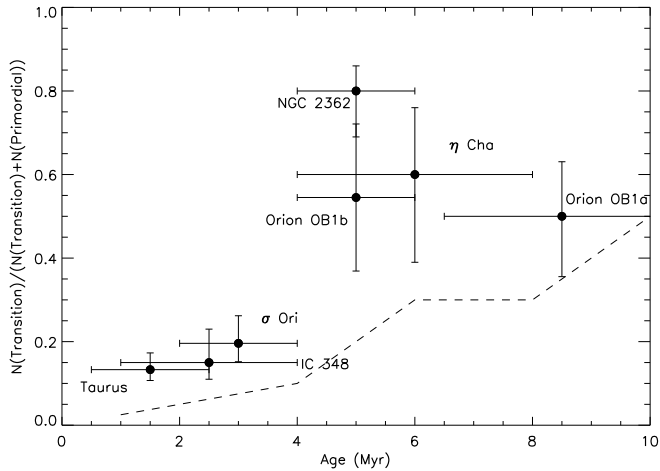}
\caption{Transitional disk frequency vs. time for 0.5--1.4 M$_{\odot}$ 
stars compared to predictions for a 0.5 Myr transition timescale from 
\citet[][dashed line]{AlexanderArmitage2009}.  To determine 
the appropriate spectral type range for each cluster, we adopt the \citet{Baraffe1998}
 isochrones and the \citet{Currie2010} T$_{e}$ scale.}
\end{figure}
As shown by Figure 2, the transitional disk frequency increases with time, climbing above $\sim$ 40--50\% for all clusters 
older than 4 Myr.  For reference, I overplot the approximate locus of transitional disk frequencies assuming a clearing 
timescale of $\sim$ 0.5 Myr \citep{AlexanderArmitage2009}.   Transitional disk frequencies for 
all clusters, most obviously those older than 3 Myr, lie above the 0.5 Myr locus indicating that the 
typical transitional disk lifetime must be longer.   \citet{Muzerolle2010} also find that the transitional disk frequency 
increases with time.  Their frequencies for 3--10 Myr-old clusters agree with those here if 
both disks with inner holes and homologously depleted ("warm"/"weak" excess sources in their terminology) 
are counted as transitional disks.

\citet{Luhman2009} dispute these claims, arguing that many of the transitional disks identified by 
\citet{CurrieLada2009} and \citet{SiciliaAguilar2009}, especially the homologously depleted disks, 
are simply primordial disks with dust settling.  To make this argument, they determine the 
IRAC and MIPS colors for a two grain population disk model from \citet{Dalessio2006} 
with an extremely low accretion rate ($\dot{M}$ = 10$^{-10}$ M$_{\sun}$ yr$^{-1}$) 
and a "depletion factor", $\epsilon$ = 0.001, which removes 99.9\% of the small dust grains and 
compare these colors to those for several 1--10 Myr-old clusters. 

Besides not knowing whether their adopted disk model is physically realistic\footnote{Numerical simulations indicate 
that micron-sized grains must produce small fragments when they grow, otherwise mid-IR emission from disks would disappear $\sim$ 100-1000 times faster than observed 
\citep{Dullemond2004}.  Since fragmentation in a collisionally dominated disk produces copious amounts of small dust, it is unclear whether disks can lose 99.9\% of 
their small dust while retaining their 10 $\mu m$ to 1 mm-sized dust, especially in the presence of turbulence.}, their analysis contains  
questionable features which undermine their conclusions.  For example, the Luhman et al. conclusions are extremely sensitive to 
 assumed values for their free parameters ($\epsilon$ and $\dot{M}$).  Assuming either that disks only remove 99\% of their small dust instead of 99.9\% or have a still-tiny accretion rate of $\dot{M}$ = 10$^{-9}$ 
M$_{\odot}$ yr$^{-1}$ revises their fiducial IRAC and MIPS primordial disk colors redwards by $\sim$ 0.5 mags, which then yields the results from \citet{CurrieLada2009}.  More generally, 
of the 16 combinations of $\epsilon$ and $\dot{M}$ shown in Figure 13 of D'Alessio et al., 
the only combination yielding the Luhman et al. results is the one Luhman et al adopts.  
Thus, models yielding the conclusions of \citet{Luhman2009} occupy a very narrow range of parameter space.  Many 1--10 Myr-old sources 
identified as transitional by Currie et al. and others are accreting at rates that must be much greater than 10$^{-10}$ M$_{\odot}$ yr$^{-1}$ (e.g. ID-36 
in NGC 2362), so the appropriate D'Alessio et al model in many cases clearly cannot be the one they adopt.

Furthermore, SED modeling of sources show that some claimed to be primordial by 
\citeauthor{Luhman2009} such as ID-3 in NGC 2362 have large inner holes.  
The \citeauthor{Luhman2009} classification scheme would also identify UX Tau and LkCa 15 as primordial disks.  
However, as the same authors showed in previous work \citep[e.g.][]{Espaillat2008}, these disks have large, $\sim$ 50 AU-scale 
holes consistent with substantial disk clearing and inconsistent with a primordial disk morphology.
Using the colors of UX Tau, LkCa 15, DM Tau and others as an empirical division between primordial and transitional disks 
also recovers the results from \citet{CurrieLada2009}\footnote{Additionally, \citet{Luhman2009} 
mischaracterize the \citet{CurrieLada2009} criteria as identifying transitional disks with emission weaker than the median Taurus SED.
Rather, the \citet{CurrieLada2009} selection criteria is a dual one, relying on comparing observed SEDs to the \textit{lower quartile} 
Taurus SED and the grid of disk models from \citet{Robitaille2006}.  This criteria was clarified in \citet{CurrieKenyon2009}.  
Several NGC 2362 source are discarded by \citet{Luhman2009} who claim, based on inspection of the post-BCD mosaic they 
lack have SNR $<$ 2--3.  However, PRF-fitting photometry and supplemental analysis of the processed mosaic show 
that most have SNR $>$ 5 and thus should not be excluded.}.
\newpage
\begin{figure}[ht]
\centering
\plotone{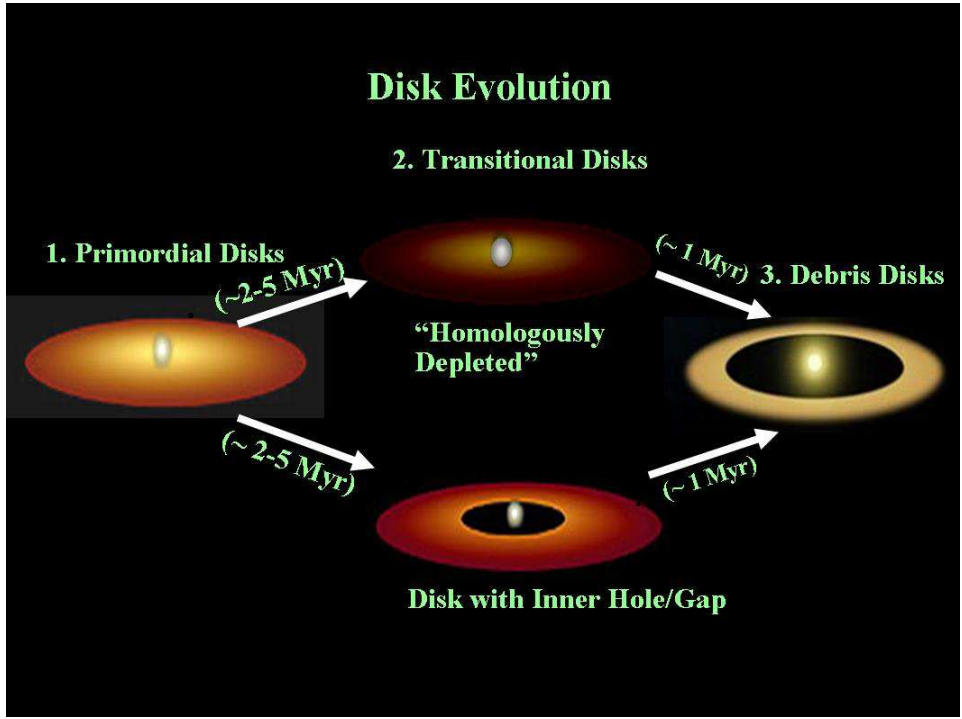}
\caption{A schematic illustrating disk evolution from the primordial disk phase to the 
debris disk phase, including the morphologies and lifetimes of transitional disks 
\citep{CurrieLada2009}.}
\label{diskevo}
\end{figure}
Figure \ref{diskevo} summarizes the general picture of disk evolution outlined by 
\citet{CurrieLada2009}.  Most optically-thick 
primordial disks last between 2 Myr and 5 Myr, depending on stellar mass.  After this time, they begin 
to show evidence of dust clearing either at a wide range of disk locations simultaneously (a "homologously 
depleted" disk) or open a gap/inner hole and deplete from the inside out.   After 5 (3, 10) Myr, 
most solar-mass stars (2.0 M$_{\odot}$ stars, subsolar-mass stars) either lack disks or have debris disks 
\citep[e.g][]{Carpenter2009,Hernandez2009,Currie2008}.

\section{Mechanisms Responsible for Transitional Disks}
Comparing transitional disk properties to other disk diagnostics and stellar properties constrains 
plausible mechanisms responsible for forming such disks.  
  For transitional disks with inner holes, the hole size is 
correlated with stellar mass/T$_{e}$ and x-ray luminosity but is only weakly correlated/not correlated 
with accretion rate \citep{Kim2009}.  \citet{Najita2007} identify a correlation between 
accretion rate and disk mass in Taurus: transitional disks systematically have larger 
 disk masses (as inferred from submm data) than primordal disks with the 
same accretion rates.  Based on a sample of stars with transitional 
disks at a wider range of ages, \citet{Cieza2008,Cieza2010} find 
instead that disk mass and accretion rate are anticorrelated.  
Since clearing induced by a gas giant planet probably occurs in 
more massive disks\footnote{Because more massive disks produce gas giant planets \citep{KennedyKenyon2008,Currie2009}.} and 
photoevaporate clearing occurs after disks have lost significant mass 
\citep[][]{Clarke2001}, the conflicting results over 
$\dot{M}$ vs. M$_{disk}$ may point to multiple processes responsible for forming inner holes 
\citep{Cieza2010,AlexanderArmitage2009}.  
Since homologously depleted transitional disks lack evidence for inner holes/gaps, their morphologies 
are not naturally explainable by gap-opening planets.
Based on results from \citet{Cieza2010} and \citet{Muzerolle2010}, their accretion properties 
and disk masses may be consistent with a simple viscous draining of disk material with time, perhaps 
accelerated by weak photoevaporation.

The homologously depleted disk morphology may occupy an important region of disk/star parameter space.
Compared to 1--3 M$_{\odot}$ stars, the frequency of massive, RV-detected planets orbiting $\sim$ 0.5 M$_{\odot}$ stars
 is far lower \citep{Johnson2007}.  These low-mass stars 
may have a higher frequency of homologously depleted transitional disks \citep[e.g.][]{Muzerolle2010}. 
Disk clearing timescales from photoevaporation 
may also be longest for the lowest-mass stars \citep[$\ge$ 1 Myr][]{Gorti2009}.  
A homologously depleted disk morphology is then a plausible outcome of a disk that fails to form 
gas giant planets and is irradiated by a low stellar UV flux.  

\section{Uncertainties and Future Work}
Many caveats qualify these conclusions about transitional disk 
properties.  Since there may be multiple mechanisms responsible for explaining the 
morphologies of transitional disks a "transition timescale" derived from the 
number of all transitional disks regardless of morphology is only a
mean value of the timescales from a number of different mechanisms.  
For example, it's entirely possible that disk clearing from gap-opening 
planets explains a subset of transitional disks and often occurs on 
 fast $\sim$ 0.1 Myr timescales, whereas other mechanisms operate more 
slowly.  Differentiating between a rapidly operating disk clearing mechanism 
and one that operates less frequently requires 
sophisticated multiwavelength analyses of disks around stars with 
a wide range of ages \citep[e.g.][]{Cieza2010}.

The disk clearing time for many candidate mechanisms (e.g. gas giant planet 
formation, UV photoevaporation) depends on properties that likely have a 
large intrinsic dispersion, such as disk viscosity, initial angular momentum, mass. 
In turn, there may be an intrinsic dispersion in transition timescales like the 
intrinsic dispersion in protoplanetary disk lifetimes.  While 
the mean transition timescale (averaged over all disk clearing mechanisms) 
may be closer to $\sim$ 1 Myr as found by \citet{CurrieLada2009}, it is 
quite plausible that many disks, including those in the youngest regions like Taurus, disperse 
on much shorter timescales \citep{CurrieKenyon2009}.

Finally, inferring gas and dust masses in transitional disks from 
IR to submm dust emission is fraught with uncertainties. 
  Deriving the dust mass from IR to submm SED modeling requires 
assuming a dust opacity, $\kappa$.  Since the dust 
opacity is not strictly known, what is actually derived from 
SED modeling is the product of the dust mass and dust opacity, not 
simply the dust mass.  The total disk mass is even more uncertain since it requires assuming a 
gas-to-dust ratio.   The models used here, in \citet{CurrieLada2009}, 
and other work \citep[e.g.][]{Cieza2010, AndrewsWilliams2005} assume 
standard values for $\kappa$ and a solar gas-to-dust ratio.  However, 
comparing derived disk masses from these methods to estimates based 
on accretion rates indicates that the former methods may systematically 
underestimate the total disk mass \citep{AndrewsWilliams2007}. 
 SED modeling and accretion diagnostics for both types of 
transitional disks indicate that they are evolutionary 
states distinct from primordial disks.  But quantitatively 
assessing how transitional disks probe clearing of both gas 
and dust requires diagnostics of circumstellar gas sensitive 
to the bulk gas mass in planet-forming regions. 

New and upcoming facilities allow studies of transitional disks that complement 
and clarify results based on Spitzer work.  The
Atacama Large Millimeter Array (ALMA) will provide extremely sensitive, high 
angular-resolution data to better constrain 
 spatial extents, inner hole sizes, surface density profiles, and 
 masses of known transitional disks.  ALMA may also identify new 
transitional disks with inner holes/gaps too small to be confidently inferred 
from SED modeling.

The most crucial, hitherto underexplored angle for investigating 
transitional disks is the disks' gas content.   Gas diagnostics for transitional disks are typically 
limited to accretion, which identifies the presence of hot circumstellar gas near 
the star.  However, gas in cooler planet-forming regions is more 
relevant in the context of disks evolving from pre planet building to post-planet building 
stages.  \textit{Herschel} offers a sensitive probe of far-IR line emission to 
identify cool gas in disks.  \textit{Herschel} programs such as GASPS will survey 
many 1--10 Myr-old stars for evidence of circumstellar gas and thus provide 
strong constraints on gas dissipation as a function of time.  Comparing the 
gas properties of transitional disks with those for primordial disks 
will more definitively determine how transitional disks 
are clearing their gas, complementing studies of dust clearing.


\acknowledgements  I thank Lucas Cieza, Cornelius Dullemond, Aurora Sicilia-Aguilar, Scott Kenyon, 
Sean Brittain, and Catherine Espaillat for helpful conversations.  
  This work was supported by a NASA Postdoctoral Fellowship.
{}
\end{document}